# Trustworthy Edge AI: Security, Privacy, and Resilience for Distributed Intelligence

Jaydip Sen[1] and Subhasis Dasgupta[2]
Department of Data Science and Artificial Intelligence
Praxis Business School, Kolkata, INDIA
email: {[1]jaydip.sen, [2]subhasisdasgupta1}@acm.org

**Abstract:** This chapter examines the principles and emerging research directions for building secure, privacy-preserving, and resilient Edge Artificial Intelligence (Edge AI) systems. As intelligence shifts from centralized clouds to distributed edge environments, challenges arise in ensuring data integrity, privacy, and reliable operation under adversarial and uncertain conditions. The chapter analyzes Edge AI architectures, emphasizing the expanded attack surface due to heterogeneous devices, resource constraints, and dynamic networks. It reviews core security mechanisms, including hardware-rooted trust, zero-trust models, secure communication, and attestation. Privacy-preserving approaches, such as federated learning, secure aggregation, and differential privacy, enable collaborative intelligence without centralized data sharing. Resilience strategies, including fault tolerance, adversarial robustness, and adaptive orchestration, are also explored. Case studies from smart cities, autonomous vehicles, healthcare, and industrial IoT illustrate practical trade-offs. The chapter advocates a holistic, cross-layer design integrating security, privacy, and resilience for trustworthy edge intelligence.

## List of Abbreviations

AI: Artificial Intelligence

ARM: Advanced RISC Machine

CNN: Convolutional Neural Network

DP: Differential Privacy

DP-SGD: Differentially Private Stochastic Gradient Descent

Edge AI: Edge Artificial Intelligence

FHE: Fully Homomorphic Encryption

FL: Federated Learning

GAN: Generative Adversarial Network

GPU: Graphics Processing Unit

HE: Homomorphic Encryption

IEC: International Electrotechnical Commission

IID: Independent and Identically Distributed

IoT: Internet of Things

IIoT: Industrial Internet of Things

IoMT: Internet of Medical Things

IP: Internet Protocol

ISO: International Organization for Standardization

LiDAR: Light Detection and Ranging

LIME: Local-Interpretable Model-agnostic Explanation

LoRaWAN: Long Range Wide Area Network

LSTM: Long Short-Term Memory

MiTM: Man-in-the-Middle

ML: Machine Learning

NIST: National Institute

OTA: Over-the-Air

RMF: Risk Management Framework

RSU: Road Side Unit

SGX: Software Guard Extensions

SHAP: SHapley Additive exPlanations

SMC: Secure Multiparty Computation

TEE: Trusted Execution Environment

TLS: Transport Layer Security

TPM: Trusted Platform Module

V2X: Vehicle-to-Everything

XAI: eXplainable Artificial Intelligence

YOLO: You Only Look Once

# 1. Introduction

The rapid growth of intelligent sensing, ubiquitous connectivity, and data-driven decision-making has transformed modern computing architectures, motivating a shift from centralized cloud-based processing to distributed Edge Artificial Intelligence (Edge AI). In this paradigm, AI models are deployed and executed close to data sources, enabling localized inference, reduced latency, and improved bandwidth efficiency [1, 2]. Advances in model compression, energy-efficient hardware, and collaborative learning have further accelerated the adoption of distributed intelligence across heterogeneous edge environments [3]. At the same time, processing data near its source supports data minimization and reduces privacy risks associated with large-scale data transmission and centralized storage [4].

Yet decentralization introduces complex trust challenges. Centralized cloud environments permit relatively controlled and homogeneous security governance. In contrast, edge ecosystems consist of geographically dispersed, heterogeneous, and resource-constrained devices often deployed in physically accessible locations. This distribution expands the attack surface and complicates lifecycle management. Trustworthiness—encompassing reliable, secure, and ethically aligned system behavior—must therefore be treated as a foundational design objective rather than an auxiliary safeguard [5].

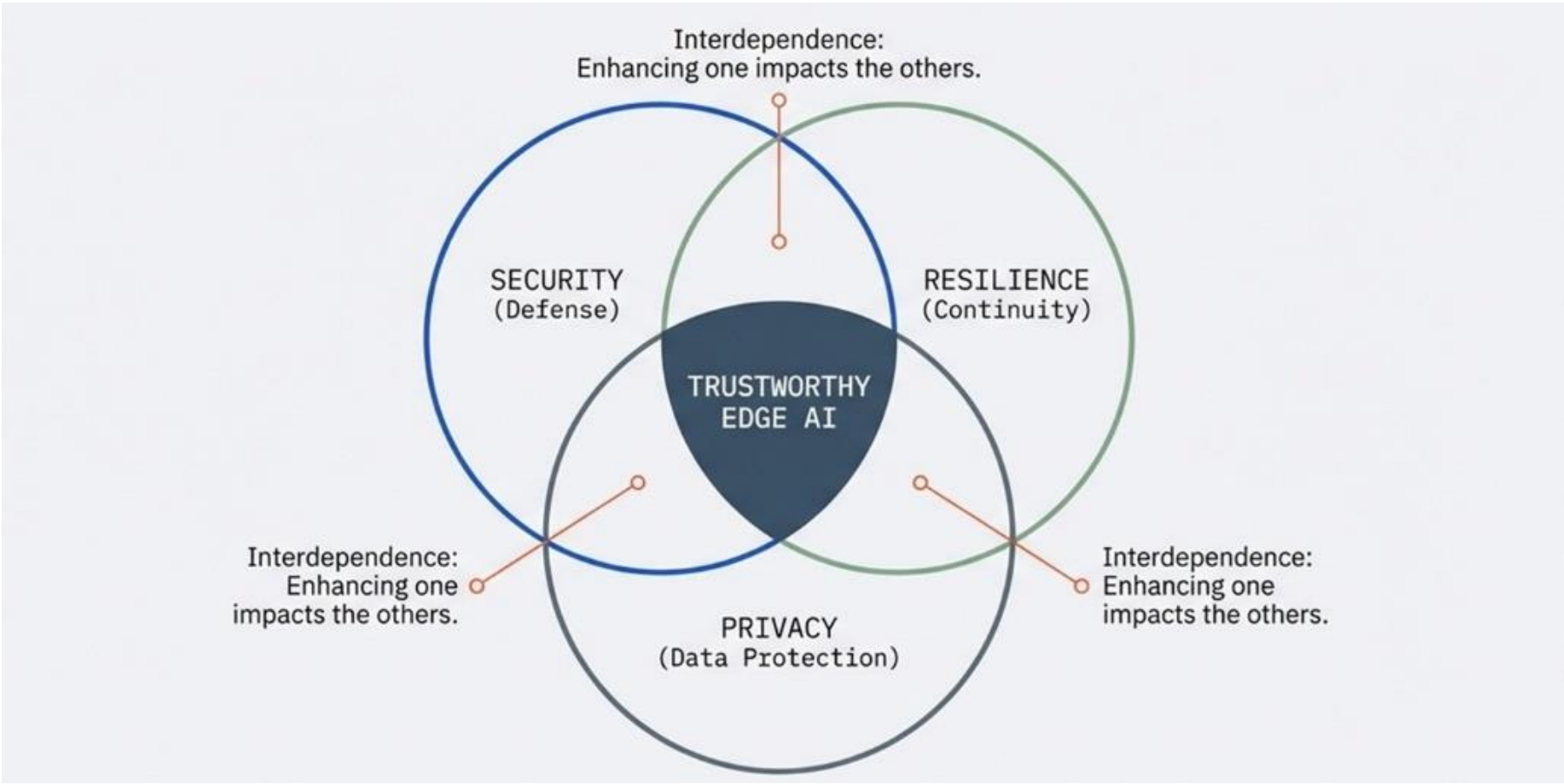

**Figure 1.** Interdependence of security, privacy, and resilience in trustworthy edge AI. Trustworthy edge intelligence emerges from the intersection of three foundational dimensions: security (defense against threat), privacy (protection of sensitive data), and resilience (continuity under failure or attack). Enhancements in one dimension influence and reinforce the others.

Within Edge AI, trustworthiness rests on the integrated realization of security, privacy, and resilience. Security addresses protection against unauthorized access, data exfiltration, model theft, firmware manipulation, and adversarial interference. Edge devices are particularly vulnerable to physical tampering and side-channel attacks due to their exposure in uncontrolled environments. AI-specific threats, including adversarial examples, model extraction, and data poisoning, further amplify risk, particularly in distributed learning frameworks where malicious participants may manipulate model updates [6, 7]. Distributed learning frameworks, including federated learning, further complicate security considerations because malicious participants may attempt to manipulate model updates or inject poisoned data [8].

As illustrated in Figure 1, trustworthy Edge AI arises from the intersection of security, privacy, and resilience, each exerting reciprocal influence on the others.

Privacy considerations remain equally critical. Although localized processing reduces raw data transmission, collaborative learning and distributed inference may leak sensitive information through gradients or model parameters. Attacks such as membership inference and model inversion demonstrate that trained models themselves can become vectors of information leakage [9]. Techniques including differential privacy, secure aggregation, homomorphic encryption, and trusted execution environments offer mitigation pathways; however, their deployment on resource-limited edge devices demands careful balancing of computational overhead, latency, and energy consumption.

Resilience complements security and privacy by emphasizing sustained functionality under disruption. While security mechanisms aim to prevent compromise, resilience mechanisms ensure continuity when prevention is incomplete. In distributed edge environments, resilience encompasses fault tolerance, redundancy, adaptive reconfiguration, anomaly detection, and self-healing coordination. Nodes may fail due to hardware faults, connectivity interruptions, or adversarial compromise. A resilient architecture must detect anomalies, isolate affected components, and maintain acceptable performance under degraded conditions. Unlike centralized systems with single control points, edge environments require decentralized coordination and dynamic trust recalibration.

The architectural characteristics of Edge AI systems and their corresponding implications for security, privacy, and resilience are summarized in Table 1.

**Table 1.** Architectural characteristics of edge AI and their implications for trustworthiness

| Architectural Feature | Description | Security Implications | Privacy Implications | Resilience Implications |
|---|---|---|---|---|
| Distributed topology | Intelligence across heterogeneous nodes | Expanded attack surface | Data locality advantages | Failure propagation risk |
| Resource constraints | Limited computing, memory, energy | Limited cryptographic overhead | On-device processing | Vulnerability to overload |
| Physical exposure | Deployment in untrusted environments | Tampering risk | Device compromise | Need for hardware attestation |
| Intermittent connectivity | Unstable network links | Authentication challenges | Reduced central logging | Need for autonomous operation |
| Heterogeneity | Diverse hardware/software stacks | Patch management complexity | Policy inconsistency | Orchestration difficulty |

These dimensions, security, privacy, and resilience, are inherently interdependent. Enhancing one without considering the others can introduce unintended trade-offs. Strong encryption may increase latency and energy usage; redundancy may strain constrained resources; privacy-preserving noise injection may degrade model accuracy. Consequently, Edge AI demands a holistic, cross-layer design philosophy in which such trade-offs are explicitly modeled and systematically managed. Trustworthiness should emerge from integrated architectural decisions spanning hardware, communication protocols, learning algorithms, and orchestration mechanisms.

Edge environments present additional structural challenges. Device heterogeneity complicates uniform policy enforcement; resource limitations restrict computationally intensive protections; physical exposure increases susceptibility to tampering; distributed learning introduces model-level attack vectors; and dynamic network topologies necessitate continuous identity verification and trust negotiation. These characteristics call for security architectures purpose-built for decentralized intelligence rather than adapted from cloud-centric assumptions.

This chapter provides a comprehensive and technically rigorous examination of how security, privacy, and resilience can be systematically embedded into Edge AI system design. It articulates the architectural foundations of distributed intelligence, analyzes emerging threat landscapes, evaluates protective mechanisms suitable for constrained devices, and advances an integrated framework that treats resilience as co-equal with security and privacy. The discussion spans theoretical foundations and practical implementation strategies, including cryptographic techniques, secure hardware primitives, adversarial robustness methods, and cross-layer trust orchestration.

## 2. Security Landscape in Edge AI Systems

The transition from centralized cloud-based AI to Edge Artificial Intelligence (Edge AI) expands not only the opportunities of distributed intelligence but also its security exposure. Edge AI systems operate in heterogeneous, geographically dispersed, and often resource-constrained environments, subjecting them to a broader spectrum of threat vectors than traditional cloud deployments. In centralized infrastructures, security controls are consolidated within protected data centers under administrative governance. By contrast, Edge AI ecosystems span multiple tiers of devices, ranging from edge servers to constrained sensors, with diverse computational capabilities, connectivity profiles, and security configurations [1, 10]. This heterogeneity produces a fragmented and dynamic security landscape in which vulnerabilities at any layer can propagate across the system.

Edge devices are frequently deployed in semi-trusted or untrusted physical environments, increasing susceptibility to direct adversarial access. Unlike cloud servers housed in hardened facilities, edge nodes positioned in industrial plants, public spaces, or remote installations may be exposed to tampering, reverse engineering, or hardware manipulation [11]. Such exposure broadens the attack surface, facilitating cryptographic key extraction, malicious firmware injection, or induced malfunction through environmental manipulation. Resource limitations further exacerbate risk: constrained processing capacity, memory, and power availability restrict the feasibility of computationally intensive security mechanisms, compelling designers to balance performance, energy efficiency, and protection strength [3].

A security concern particularly salient in distributed learning settings is *model poisoning*. In federated and collaborative training frameworks, participating edge devices compute local model updates and transmit parameters for aggregation [8]. Although this paradigm avoids raw data centralization, malicious participants may intentionally craft updates to corrupt the global model. Poisoning attacks may degrade overall accuracy or implant backdoors that trigger targeted misclassification under specific inputs [12–14]. Because aggregation servers lack direct visibility into local data distributions, distinguishing benign statistical heterogeneity from adversarial manipulation becomes inherently challenging.

Adversarial perturbations represent another critical threat. Carefully crafted, imperceptible input modifications can induce incorrect model predictions. Initially demonstrated in image classification, adversarial examples have been extended to speech, text, and time-series modalities [6, 15]. In Edge AI deployments, where models interact directly with physical sensors, adversarial manipulation may occur at the data source. Altered traffic signage, spoofed environmental readings, or manipulated industrial signals can provoke unsafe or erroneous decisions in autonomous systems [16, 17]. Limited inter-node communication and intermittent connectivity further complicate coordinated detection.

Integrity risks also arise at the data acquisition layer. IoT and edge sensors operating in exposed environments are susceptible to false data injection, packet replay, or signal spoofing [18]. Even without compromising the model itself, adversaries can distort inference outcomes by manipulating input streams. In industrial contexts, stealthy injection of falsified telemetry may mislead predictive maintenance or anomaly detection systems, leading to operational disruption or safety hazards.

Firmware compromise constitutes a persistent and often underestimated threat. Edge devices commonly execute embedded firmware that may remain unpatched for extended periods, accumulating exploitable vulnerabilities [19]. If update mechanisms or bootloaders are subverted, attackers can install malicious firmware capable of intercepting communications, altering sensor buffers, or redirecting model execution. Secure boot processes—leveraging cryptographic signature verification during startup—mitigate unauthorized software deployment [20, 21]. However, large-scale implementation across heterogeneous platforms introduces operational complexity, secure key provisioning challenges, and reliance on hardware roots of trust.

Hardware-based Trusted Execution Environments (TEEs), including ARM TrustZone and Intel SGX, provide isolated execution domains that protect sensitive computations and cryptographic material from compromise by other software layers [22, 23]. In edge deployments, TEEs can encapsulate inference engines or aggregation logic to reduce exposure. Nevertheless, enclave technologies incur performance overhead and remain vulnerable to certain side-channel attacks [24]. Additionally, many microcontroller-class devices lack comprehensive enclave support, limiting uniform adoption.

Blockchain and distributed ledger mechanisms have been proposed to enhance transparency and auditability in distributed Edge AI networks. Recording model updates, device attestations, or configuration changes on tamper-evident ledgers can strengthen accountability across nodes [25]. In federated learning, ledger-backed coordination may reduce dependence on a centralized aggregator. Yet, consensus protocols introduce latency and computational burden that may conflict with real-time constraints. Lightweight or permissioned blockchain variants mitigate some overhead but introduce trade-offs between decentralization, scalability, and efficiency.

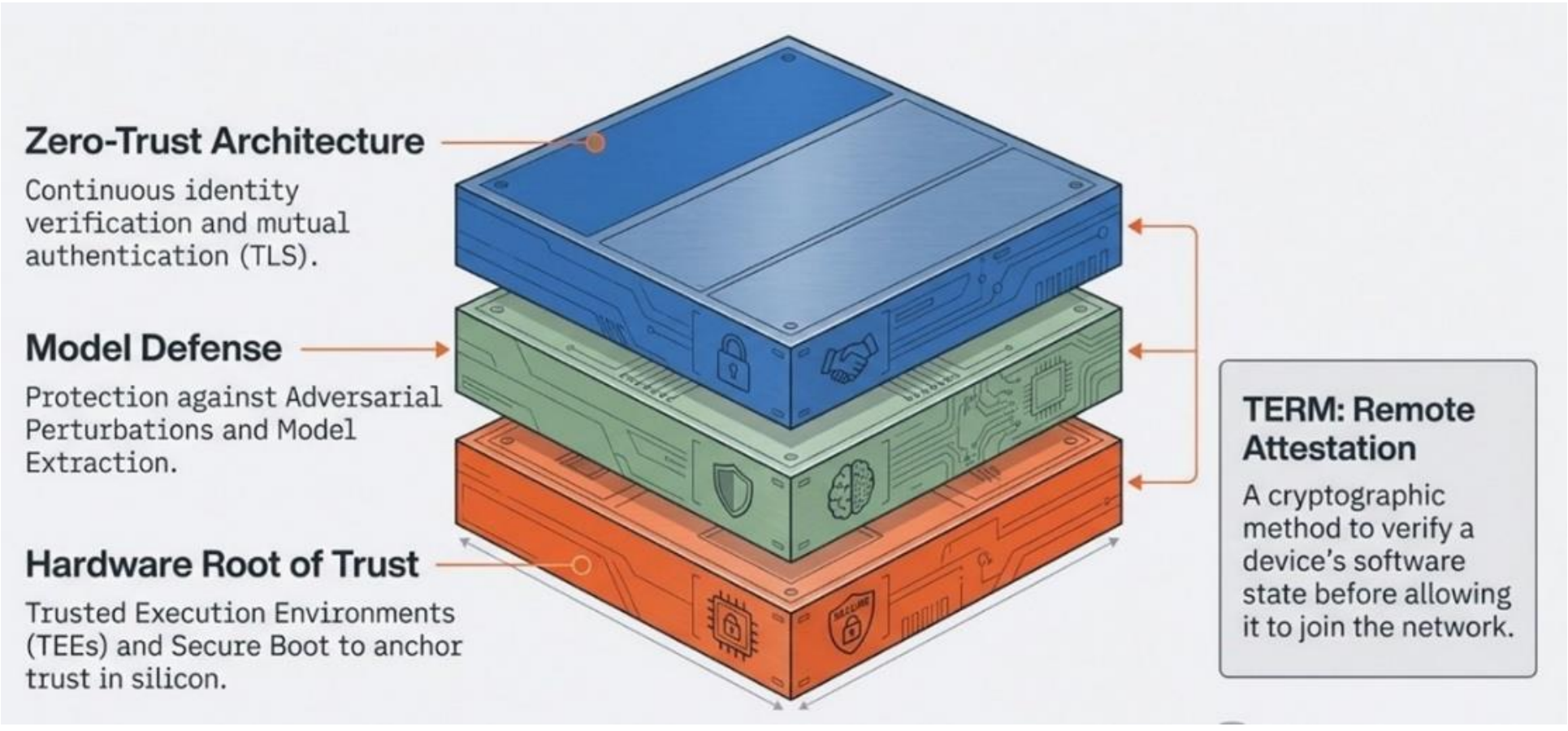

**Figure 2.** Conceptual depiction of a hardware-anchored, multi-layer security framework for Edge AI. The architecture integrates hardware root of trust, model-level defenses against adversarial manipulation and extraction, and zero-trust network principles with continuous authentication and remote attestation.

Figure 2 presents a layered security architecture for Edge AI in which trust originates at the hardware level through secure boot and trusted execution environments, is reinforced through model defense mechanisms, and is operationalized at the network layer via zero-trust enforcement and continuous identity verification. Remote attestation enables cryptographic validation of device integrity prior to participation in distributed workflows.

Model-centric robustness techniques further address adversarial manipulation. Adversarial training integrates perturbed samples into learning pipelines to improve resilience [26]. Complementary approaches, including defensive distillation, randomized smoothing, and certified robustness methods, provide empirical or mathematically bounded guarantees against defined perturbation levels [27, 28]. While these strategies enhance robustness, they frequently increase training complexity, model size, or inference latency, posing deployment challenges for constrained edge hardware.

The tension between security rigor and computational feasibility is particularly pronounced in Edge AI. Advanced cryptographic protocols, encrypted inference, and anomaly detection mechanisms demand processing capacity, memory, and energy resources that many edge nodes cannot sustain without impairing primary operational functions [1, 3, 11]. Energy consumption is especially critical for battery-powered or remote sensing platforms. Achieving sustainable security therefore requires context-aware optimization. Tiered architectures, where lightweight protections operate at sensor nodes and more intensive safeguards execute at capable gateways, offer one pragmatic solution.

Secure aggregation protocols exemplify the interplay between confidentiality and detectability. By cryptographically masking individual model updates while enabling aggregate computation, secure aggregation enhances privacy and limits exposure of local contributions [29]. However, encryption constrains the aggregator's ability to inspect updates for anomalies or malicious behavior, illustrating the broader trade-off between privacy preservation and security transparency.

Hardware roots of trust, including Trusted Platform Modules (TPMs) and secure elements, anchor cryptographic identities and enable secure key storage and remote attestation [30]. Remote attestation allows verification of a device's software state prior to network participation. Nevertheless, attestation procedures introduce communication overhead and rely on robust key management infrastructures. Given the diversity of edge hardware, from powerful gateways to minimal sensors, universal deployment remains operationally complex.

Secure lifecycle management is equally essential. Firmware updates must be authenticated, integrity-verified, and securely delivered to prevent malicious modification. Over-the-air (OTA) update channels themselves may become attack vectors if authentication and encryption are inadequate. Compromised update servers or intercepted distributions can propagate malicious code widely, making robust update governance foundational for sustained security.

To address multilayered threats, adaptive security architectures have been proposed. These frameworks dynamically allocate protective mechanisms according to device capability, threat context, and operational priority. High-capacity edge servers may implement comprehensive defenses, including TEEs, strong cryptography, and ledger-backed auditing, while lightweight sensors employ minimal primitives and delegate complex protections upstream. Such hierarchical designs optimize resource utilization while preserving baseline safeguards.

**Table 2.** Threat landscape and mitigation strategies in edge AI systems

| Threat Category | Example Attacks | Target Layer | Potential Impact | Mitigation Strategies |
|---|---|---|---|---|
| Physical attacks | Device tampering, side-channel attacks | Hardware | Key extraction | Secure enclave, TPM |
| Network attacks | MTM, replay, spoofing | Communication | Data manipulation | TLS, mutual authentication |
| Model attacks | Model extraction, inversion | ML layer | IP leakage | Secure inference protocols |
| Data poisoning | Malicious client updates | Federated learning | Model degradation | Byzantine-resilient aggregation |
| Adversarial inputs | Perturbed inputs | Inference stage | Misclassification | Adversarial training |

In summary, the security landscape of Edge AI is defined by decentralization, heterogeneity, and constrained resources. Threats span model poisoning, adversarial manipulation, firmware compromise, data tampering, and hardware exploitation. Corresponding defenses, including trusted execution, secure boot, ledger-based integrity validation, robust training, secure aggregation, and hardware roots of trust, each entail trade-offs in complexity, latency, and energy consumption. Secure Edge AI design therefore requires a cross-layer, context-aware strategy that harmonizes protection strength with operational feasibility to sustain trustworthy distributed intelligence.

The major threat vectors in Edge AI systems and their corresponding mitigation strategies are systematically summarized in Table 2.

## 3. Privacy Preservation in Edge Intelligence

The proliferation of edge intelligence has fundamentally reshaped how data is collected, processed, and operationalized in distributed environments. Unlike centralized cloud-centric architectures, Edge AI systems operate across heterogeneous devices, ranging from smartphones and wearables to industrial controllers and vehicular nodes, where sensitive data is generated and frequently retained locally. Although this architectural shift reduces raw data transmission and mitigates certain centralized privacy risks, it simultaneously introduces new and less visible vulnerabilities. Privacy preservation in Edge AI is therefore not simply a matter of limiting data exposure; it demands a systematic rethinking of how learning algorithms, communication protocols, and cryptographic mechanisms interact under stringent resource constraints.

At the core of privacy-aware edge intelligence lies federated learning (FL), a paradigm that enables decentralized model training without directly sharing raw data [8]. In FL, participating devices compute local model updates and transmit parameter gradients or weight differentials to an aggregation server. This mechanism was initially proposed to protect user data in distributed settings such as mobile keyboards and recommendation systems [8]. However, the assumption that withholding raw data inherently guarantees privacy has proven overly optimistic. Subsequent research has demonstrated that gradient updates themselves may encode sensitive information, rendering decentralized training frameworks susceptible to inference attacks.

### 3.1 Privacy Risks in Federated and Decentralized Learning

Although federated learning reduces explicit data sharing, its iterative exchange of model parameters creates a substantial attack surface. Adversaries observing gradient updates - through a compromised server or through colluding clients - can reconstruct private inputs with remarkable fidelity. Gradient leakage attacks have shown that training images, text sequences, and fine-grained user attributes can be recovered directly from shared gradients [31, 32]. Zhu et al. demonstrated that deep learning gradients are sufficient to reconstruct original training samples via optimization-based inversion techniques [31]. Subsequent studies refined these attacks to recover high-resolution images and multi-sample batches with improved efficiency [32].

Model inversion attacks constitute another significant threat in distributed intelligence systems. In this setting, adversaries exploit access to trained models - either in white-box or black-box form - to infer sensitive features of the training data [33]. For example, inversion techniques have reconstructed facial attributes from biometric classifiers [33]. In the edge deployments, where models are distributed across numerous devices, exposure of local or global models amplifies the risk of such inference attacks.

Membership inference attacks further undermine privacy assumptions in decentralized training [9]. In these attacks, adversaries attempt to determine whether a specific data record was included in the training dataset. This threat is particularly acute in healthcare, finance, and smart infrastructure applications, where mere participation in a dataset may reveal sensitive information. Shokri et al. first formalized membership inference in machine learning models [9], and subsequent studies demonstrated that federated learning can exacerbate membership leakage due to overfitting and heterogeneous data distributions [34].

Edge deployments also introduce system-level privacy concerns. Communication patterns, timing information, and update frequency may leak metadata about user behavior, device activity, or environmental context. Side-channel signals, including power consumption, memory access patterns, and communication latency, can be exploited to infer sensitive attributes, particularly in constrained IoT devices [35]. Consequently, privacy risks in Edge AI extend beyond algorithmic leakage to encompass hardware- and network-level vulnerabilities.

## 3.2 Differential Privacy in Edge AI

Differential Privacy (DP) is a rigorous mathematical framework that ensures the output of a computation or analysis does not reveal sensitive information about any individual data record, by guaranteeing that the inclusion or exclusion of a single data point has a negligible impact on the overall result [36]. DP has emerged as one of the most rigorous mathematical frameworks for quantifying and mitigating privacy leakage in machine learning systems. A mechanism satisfies differential privacy if the inclusion or exclusion of a single data point does not significantly alter the output distribution, thereby bounding the information that can be inferred about individual records [36].

In federated learning, differential privacy is typically implemented through gradient perturbation. Noise calibrated to a predefined privacy budget ($\varepsilon$) is added either at the client level before transmission (local DP) or at the aggregation server after collecting updates (central DP) [37]. Abadi et al. introduced differentially private stochastic gradient descent (DP-SGD), which applies gradient clipping and Gaussian noise to bound sensitivity [37]. This method has since been adapted to federated environments to provide user-level privacy guarantees.

However, integrating differential privacy in Edge AI introduces nontrivial trade-offs. Excessive noise injection may degrade convergence stability and predictive accuracy, particularly under non-independent and identically distributed (non-IID) data – a common characteristic of edge devices [38]. Moreover, gradient clipping and noise generation impose computational overhead that may strain battery-powered devices. Recent studies have proposed adaptive privacy budgets that dynamically adjust noise levels based on training stage, data sensitivity, or model stability [39]. Such adaptive DP mechanisms aim to preserve accuracy while maintaining formal privacy guarantees.

Advanced variants, such as Rényi Differential Privacy (RDP) and Zero-Concentrated Differential Privacy (zCDP) are improved ways of measuring privacy loss more precisely than standard differential privacy [40]. Instead of using a single worst-case bound, RDP tracks privacy loss across different "orders" of divergence, which allows it to accumulate the effect of multiple iterations (such as repeated model updates) in a much tighter and less conservative way. Similarly, zCDP summarizes privacy loss using a smooth, well-behaved bound that makes it easier to combine privacy guarantees over many steps. Because both methods keep a more accurate account of how privacy loss builds up, they avoid overestimating the total loss that typically happens with conventional DP. This means that in iterative algorithms, less noise needs to be added at each step, leading to better model performance while still maintaining strong privacy guarantees.

## 3.3 Homomorphic Encryption for Encrypted Inference and Training

Homomorphic encryption (HE) enables computation directly on encrypted data, thereby eliminating the need to decrypt sensitive information during processing [41]. In the context of Edge AI, HE can protect both model updates and inference inputs, allowing servers to perform aggregation or prediction without accessing plaintext data.

Fully homomorphic encryption (FHE), first realized by Gentry [41], supports arbitrary computation on ciphertexts but remains computationally intensive. While recent optimizations have significantly reduced overhead [42], FHE operations still incur substantial latency and memory costs, making them challenging for resource-constrained edge devices. As a result, practical deployments often rely on partially or somewhat homomorphic schemes that support limited arithmetic operations suitable for neural network inference [43].

Encrypted inference frameworks have demonstrated that convolutional neural networks can operate over encrypted inputs with acceptable accuracy degradation [44]. Nevertheless, the computational burden can increase inference latency by orders of magnitude compared to plaintext execution. In latency-sensitive edge scenarios, such as autonomous vehicles or industrial control, such overhead may be unacceptable.

To mitigate these constraints, hybrid approaches combine homomorphic encryption with model compression or quantization techniques [45]. By reducing model complexity prior to encryption, computational costs can be partially mitigated. Additionally, hardware acceleration using *trusted* *execution environments* (TEEs) or cryptographic co-processors can improve performance, though such solutions introduce their own trust assumptions.

## 3.4 Secure Multiparty Computation in Collaborative Edge Learning

Secure multiparty computation (SMC) enables multiple parties to jointly compute a function without revealing their individual inputs [46]. In federated settings, SMC is commonly used for secure aggregation, ensuring that the central server can only access aggregated gradients rather than individual updates.

Bonawitz et al. introduced a practical secure aggregation protocol tailored for federated learning, demonstrating scalability to millions of devices [29]. The protocol relies on cryptographic masking and secret sharing to protect individual updates even if some participants drop out. Subsequent research has improved robustness against collusion and Byzantine behavior [47].

While SMC provides strong confidentiality guarantees, it introduces communication and computational overhead proportional to the number of participants. In large-scale Edge AI systems with intermittent connectivity, maintaining synchronized cryptographic protocols can be challenging. Recent work explores lightweight SMC variants optimized for bandwidth-limited IoT environments [48, 49], reducing communication rounds and employing efficient secret-sharing schemes.

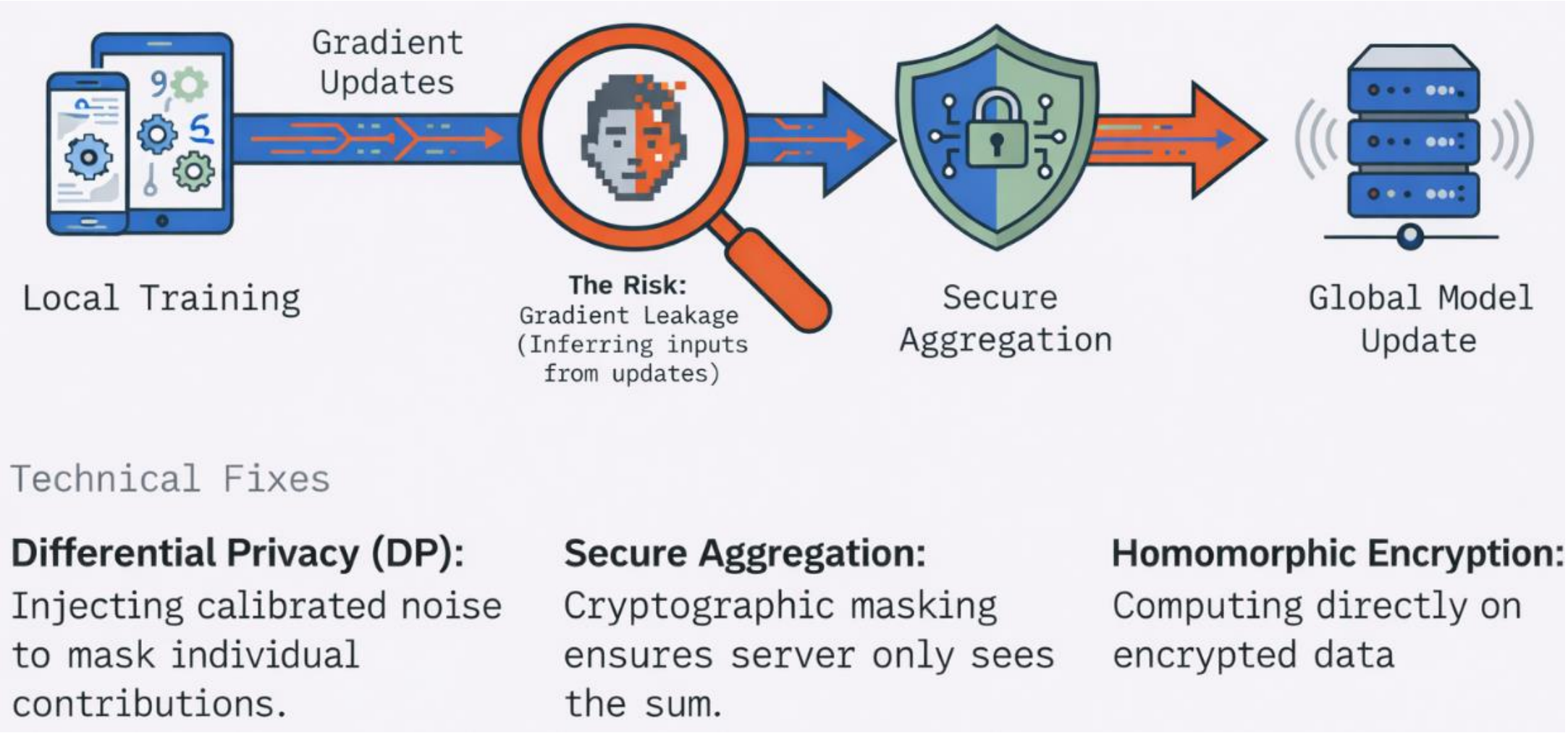

**Figure 3.** Federated learning pipeline highlighting gradient leakage vulnerabilities and corresponding privacy-preserving countermeasures. Secure aggregation prevents exposure of individual model updates, differential privacy injects calibrated noise to ensure formal privacy guarantees, and homomorphic encryption enables computation directly on encrypted parameters.

The privacy risks and mitigation mechanisms in federated learning are summarized in Figure 3.

### 3.5 Membership Inference, Model Inversion, and Emerging Attacks

Beyond classical gradient leakage, emerging attack vectors continue to challenge privacy preservation in edge intelligence. Generative Adversarial Networks (GANs) have been leveraged to strengthen inversion attacks by approximating training data distributions [50]. Adaptive adversaries may also exploit update sparsity or quantization artifacts to improve reconstruction fidelity.

Recent studies indicate that personalization techniques in federated learning - where global models are fine-tuned locally - may inadvertently increase privacy leakage by embedding user-specific signatures [51]. Cross-device correlation attacks can further exploit similarities among user updates to infer demographic or behavioral traits.

Defense strategies increasingly integrate multiple mechanisms. For example, combining differential privacy with secure aggregation reduces exposure even if aggregation is compromised [52, 53]. Encrypted gradient transmission mitigates interception risks, though it does not prevent inference from decrypted updates at the server.

### 3.6 Privacy-Utility-Efficiency Trade-offs

A central challenge in privacy-preserving Edge AI is balancing privacy guarantees with model utility and system efficiency. Strong protections, such as high-noise differential privacy or fully homomorphic encryption, may significantly degrade accuracy or increase latency. Conversely, lightweight protections may leave systems vulnerable to sophisticated attacks.

The privacy–utility trade-off is often quantified by measuring accuracy degradation as a function of privacy budget $\varepsilon$ [37]. In edge environments, however, an additional dimension - energy consumption - must be considered. Energy-aware privacy mechanisms adapt noise injection or encryption depth based on device battery level or workload [54-56], aiming to preserve operational longevity while maintaining acceptable performance.

Recent research explores multi-objective optimization to balance communication cost, model performance, fairness, and robustness in federated learning systems [57-60]. Reinforcement learning-based controllers have been explored to dynamically allocate privacy budgets across training rounds [61], enabling adaptive privacy management under evolving network and workload conditions.

### 3.7 Toward Adaptive and Context-Aware Privacy Mechanisms

Static privacy configurations are often inadequate for heterogeneous edge ecosystems. Devices differ significantly in processing capability, connectivity, and data sensitivity. Context-aware privacy frameworks therefore adjust protection mechanisms according to environmental risk, application criticality, and user preferences.

For example, healthcare edge devices may enforce stricter privacy budgets than industrial sensors collecting non-personal data. Adaptive cryptographic layering can escalate protection during high-risk communication periods or upon anomaly detection [62-64]. Hierarchical federated architectures may also partition aggregation across trusted subdomains to reduce exposure.

Emerging techniques, including split learning [65-68], privacy-constrained knowledge distillation [69-74], and confidential computing hardware extensions [22], further expand the design space. These approaches distribute computation to limit exposure of raw data and intermediate representations.

## 3.8 Synthesis and Future Directions

Privacy preservation in Edge AI is a multidimensional challenge spanning algorithmic, cryptographic, architectural, and hardware domains. Federated learning provides a foundation for decentralized training but introduces inference vulnerabilities that necessitate complementary safeguards. Differential privacy offers formal guarantees but requires careful calibration to sustain model utility. Homomorphic encryption and secure multiparty computation ensure strong confidentiality but face scalability and performance constraints. Adaptive and context-aware mechanisms represent a promising frontier, enabling dynamic balancing of privacy, utility, and resource efficiency.

As Edge AI expands into smart cities, autonomous systems, and critical infrastructure, privacy preservation must evolve from an auxiliary feature to a core design principle. Integrating mathematically rigorous privacy frameworks with lightweight cryptographic protocols and adaptive control strategies will shape the next generation of trustworthy distributed intelligence systems.

A comparative evaluation of privacy-preserving learning techniques for Edge AI, including their trade-offs in accuracy, computational overhead, and privacy guarantees, is presented in Table 3.

**Table 3**. Comparative analysis of privacy-preserving learning techniques for edge AI

| Technique | Core Principle | Advantages | Limitations | Suitable Use Cases |
|---|---|---|---|---|
| Federated Learning | Decentralized model training | Data locality | Communication overhead | Healthcare, mobile devices |
| Secure Aggregation | Encrypted parameter sharing | Confidential updates | Computational cost | Multi-organization training |
| Differential Privacy | Noise injection | Formal privacy guarantees | Accuracy trade-off | Regulatory-sensitive domains |
| Trusted Execution Environments | Hardware isolation | Strong data protection | Hardware dependency | Industrial IoT |
| Homomorphic Encryption | Computation on encrypted data | End-to-end privacy | High latency | High-value sensitive analytics |

# 4. Resilience and Dependability in Edge AI Systems

As Edge AI matures and integrates into mission-critical infrastructures, such as autonomous vehicles, industrial automation, healthcare monitoring, and smart cities, *resilience and dependability* have emerged as foundational system requirements. For distributed AI systems operating at the edge, resilience denotes the capacity to sustain acceptable levels of operational performance in the presence of faults, failures, adversarial disruptions, environmental uncertainty, and fluctuating resource availability. Unlike centralized systems, where computation and decision-making occur within controlled infrastructures, Edge AI systems comprise heterogeneous devices deployed across dynamic and often unreliable networks. These characteristics introduce challenges that conventional fault tolerance alone cannot adequately address.

This section formalizes resilience in the context of distributed AI, distinguishes it from traditional reliability constructs, and examines architectural and algorithmic mechanisms that enable resilient operation. Redundancy strategies, anomaly detection, self-healing architectures, robust optimization, and adaptive control are then examined, with illustrative case examples demonstrating how systems sustain functionality despite partial failures or malicious interference.

## 4.1 Defining Resilience in Distributed AI Systems

In distributed AI systems, resilience refers to the ability to maintain critical functionality and bounded performance degradation under both accidental faults and deliberate adversarial actions [75-77]. Resilience extends beyond traditional notions of reliability and fault tolerance: whereas reliability emphasizes minimizing failure probability and fault tolerance addresses predefined failure modes, resilience incorporates adaptive, anticipatory, and recovery capabilities that enable sustained operation under uncertain and evolving conditions.

A widely cited definition frames resilience as *"the ability of a system to absorb disturbances, recover from failures, and adapt to changing environmental and operational conditions without unacceptable degradation of performance"* [75]. For Edge AI, this definition must further account for:

- Heterogeneous hardware and software stacks, including microcontrollers, sensors, gateways, and edge servers, each with distinct reliability and resource constraints.
- Dynamic network topologies, where nodes may intermittently join, leave, or lose connectivity due to mobility or environmental disruption.
- Data variability and workload volatility, where data inputs and resource demands shift over time.
- Adversarial conditions, targeting system components, communication channels, or machine learning models themselves.

Resilience thus emerges as a systemic property arising from the coordinated orchestration of redundancy, monitoring, adaptation, control, and learning. It integrates classical fault tolerance with predictive diagnostics, adaptive optimization, and self-learning mechanisms to address both known and unforeseen threats [78-80].

## 4.2 Distinguishing Resilience from Fault Tolerance

Fault tolerance is a mature discipline in distributed computing that ensures continued operation despite specified component failures (e.g., hardware faults, software bugs, or transient errors). Techniques such as replication, checkpointing, and consensus protocols exemplify this approach [81-84]. However, traditional fault tolerance typically assumes known failure modes and emphasizes maintaining service continuity through relatively static recovery mechanisms.

Resilience, by contrast, constitutes a broader and more dynamic framework that:

- Anticipates a wider spectrum of disruptions, including adversarial manipulation and unforeseen environmental variation.
- Integrates detection, response, and adaptation loops, rather than relying solely on predefined remediation strategies.
- Enables learning over time, allowing the system to strengthen its defenses by leveraging historical disruptions.
- Accepts bounded performance degradation and prioritizes graceful degradation, instead of binary success-failure outcomes [75, 85, 86].

In Edge AI, resilience must address not only classical system faults (e.g., node crashes or communication loss) but also machine learning-specific anomalies, such as concept drift, model poisoning, and adversarial evasion. Consequently, traditional fault-tolerance mechanisms are necessary but insufficient; they must be complemented by adaptive intelligence capable of interpreting and mitigating semantic failures that transcend hardware or software defects.

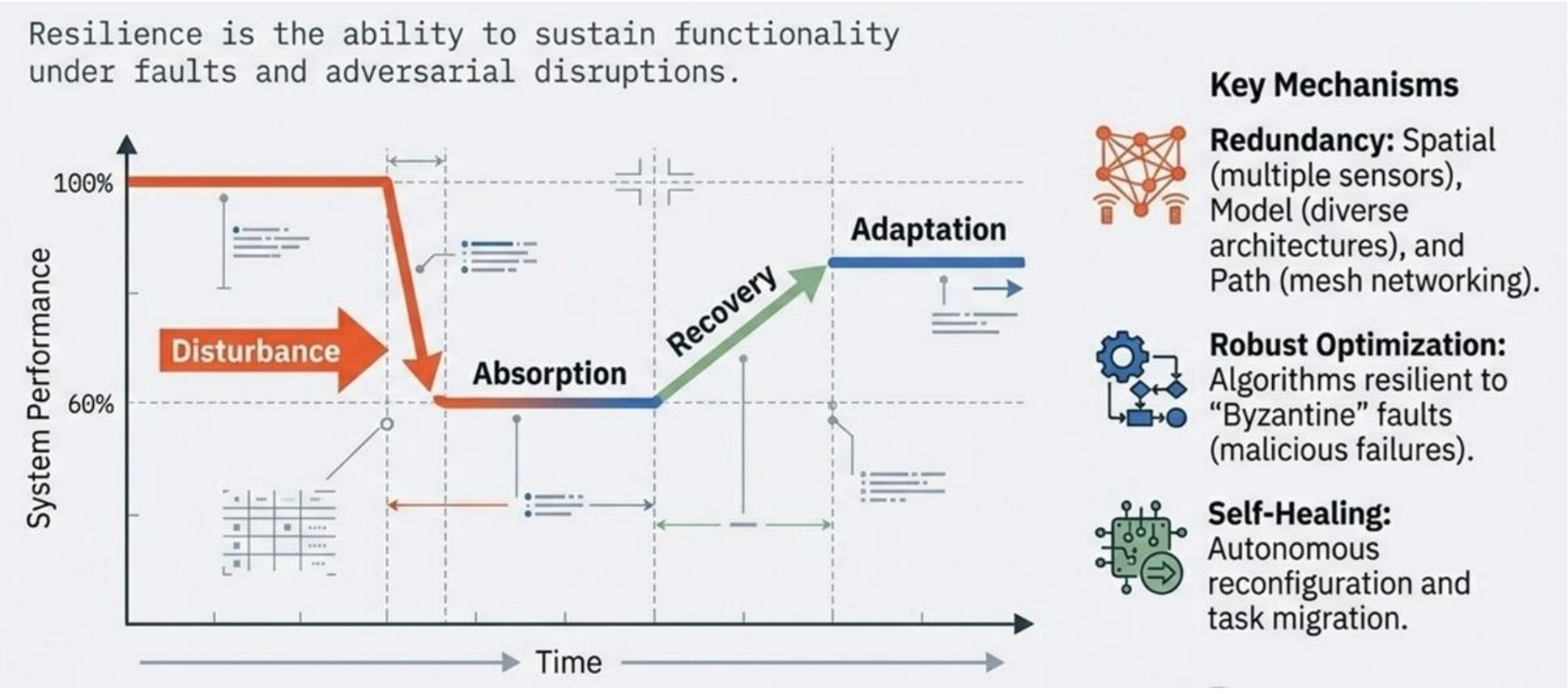

**Figure 4.** Illustration of system performance over time during a disturbance event. Following a disruption, resilient systems absorb performance degradation, recover through corrective mechanisms, and adapt to restore or enhance operational capacity. Key enabling mechanisms include redundancy (spatial, model, and path), Byzantine-resilient optimization, and autonomous self-healing.

Figure 4 depicts the performance-time profile of a resilient Edge AI systems under adversarial or fault-induced disturbances. The resilience process encompasses absorption of impact, recovery through corrective mechanisms, and adaptive reconfiguration enabled by redundancy, robust optimization, and self-healing architectures.

## 4.3 Mechanisms for Resilience in Edge AI

Resilient Edge AI systems employ mechanisms operating across multiple layers of the hardware-software stack. These mechanisms include redundancy, anomaly detection, self-healing, and robust optimization, collectively enabling sustained functionality under adverse conditions.

### 4.3.1 Redundancy Strategies

Redundancy entails duplicating critical components, communication pathways, or computational processes so that failures in one element can be compensated by another. In distributed AI systems, redundancy manifests at multiple levels:

- *Spatial redundancy:* Deploying multiple sensors or nodes covering the same region so that corrupted or failed readings can be cross validated [87].
- *Model redundancy:* Maintaining multiple model instances – potentially trained using diverse architectures or data subsets – to cross-check outputs and mitigate poisoning or corruption [88].
- *Path redundancy:* Establishing alternative communication routes to prevent network partitioning from isolating subsystems [89].

In autonomous driving fleets, for example, redundant distance and vision sensors enable environmental inference even when individual sensors fail or are compromised, thereby preserving operational safety [90-92].

### 4.3.2 Anomaly Detection

Anomaly detection identifies patterns that deviate from established baselines. In Edge AI, anomalies may stem from sensor faults, communication interference, adversarial inputs, or internal model degradation.

Detection mechanisms operate at multiple levels:

- *Local anomaly detection:* Executed on individual devices to identify abnormal sensor readings or anomalous resource behavior (e.g., autoencoder-based detection of out-of-distribution inputs) [93–95].
- *Global anomaly detection:* Aggregating cross-device signals to detect systemic irregularities, such as correlated drift across nodes, potentially indicating model corruption [96].

Real-time anomaly detection is especially critical in industrial IoT and safety-critical infrastructures, where early detection can prevent cascading failures.

### 4.3.3 Self-Healing Mechanisms

Self-healing denotes the system's capacity to autonomously recover from detected faults without external intervention. Common strategies include:

- *Reconfiguration:* Migrating tasks from compromised nodes to healthy ones.
- *Reinitialization:* Resetting malfunctioning components or retraining degraded models.
- *Rollback:* Restoring prior system states from checkpoints or snapshots [97-99].

In Edge AI environments, self-healing may involve retraining local models using updated data after performance degradation is detected or redistributing workloads to prevent overload-induced failures.

### 4.3.4 Robust Optimization

Robust optimization seeks to ensure reliable performance under uncertainty, perturbation, or partial data loss. In Edge AI, this includes:

- *Robust federated learning:* Mitigating malicious or unreliable updates throughout trust-weighted aggregation or statistical outlier filtering [47,100-102].
- *Adversarially robust training:* Incorporating perturbed samples during training to enhance robustness against adversarial inputs [27, 103-105].
- *Resilient aggregation rules:* Employing median or trimmed mean aggregation to reduce the impact of corrupted gradients [106-109].

Robust optimization closely aligns with resilience objectives by explicitly designing for degraded or adversarial conditions.

## 4.4 Adaptive Control and Feedback in Resilient Edge AI

Adaptive control mechanisms dynamically adjust system behavior in response to environmental variation and performance feedback. In resilient Edge AI systems, adaptive controllers:

- Monitor performance metrics, such as accuracy, latency, and resource utilization.
- Dynamically tune parameters, including sampling rates, model complexity, and resource allocation.
- Trigger fallback mechanisms when anomalies threaten operational continuity [110-112].

For instance, in smart grid edge frameworks, controllers adjust load balancing and model execution priorities in real-time based on demand fluctuations and network conditions, maintaining stability during peak loads or partial outages [113, 114].

## 4.5 Case Examples of Resilient Architectures

To contextualize the preceding discussion on resilience mechanisms, it is instructive to examine how these principles are realized in practical Edge AI deployments. The following case examples illustrate how architectural strategies such as redundancy, distributed anomaly detection, and adaptive coordination are implemented across real-world systems to ensure continuity and robustness under failure and adversarial conditions.

### 4.5.1 Industrial IoT Infrastructure

In industrial settings, distributed sensors and controllers monitor machinery, environmental conditions, and process states [115]. A resilience framework integrates sensor redundancy, anomaly detection models at the edge, and robust optimization in model aggregation. During unanticipated sensor failures, redundant sensors and cross-device anomaly checks ensure that critical control decisions can still be made. Advanced systems use predictive diagnostics to anticipate failure before it occurs, enabling preemptive maintenance [116-118].

An example of hierarchical redundancy and layered anomaly monitoring is seen in adaptive anomaly detection schemes designed for *hierarchical edge computing* where IoT devices, regional edge servers, and centralized nodes cooperate to balance resource use and detection accuracy. These layered structures allow local, resource-light detectors at the device level and more sophisticated analyses at higher levels, enabling continued monitoring and adaptive responses even when individual nodes underperform or fail [119, 120].

### 4.5.2 Autonomous Vehicle Networks

Autonomous driving systems rely on a fusion of perception, localization, and decision models. These systems must remain operational under partial sensor loss, communication interruptions, or adversarial signals. Multi-sensor redundancy (e.g., combining light detection and ranging (LiDAR), radar, and cameras) is combined with robust perception models that can tolerate missing or corrupted inputs. When a sensor fails, the system reweights inputs and continues safe navigation. Recent architectures also incorporate anomaly detectors that identify out-of-distribution events and trigger safe fallback behaviors (e.g., slow down or stop) [121, 122].

### 4.5.3 Smart Healthcare Monitoring

Wearable devices and edge gateways conduct continuous monitoring of patient vitals. Resilience in this context demands self-healing and adaptive sampling to cope with network outages or battery depletion. Adaptive control algorithms monitor device battery health and adjust data sampling rates to prolong monitoring without losing critical alerts [123, 124]. Anomaly detection plays a key role in identifying sensor drift due to wear or interference.

## 4.6 Balancing Resilience with Efficiency

While resilience is crucial, achieving it often comes at a cost in terms of resource consumption or latency. Redundancy requires additional hardware and communication overhead. Robust optimization and anomaly detection consume computational resources. Adaptive control mechanisms require continuous monitoring and potentially frequent reconfiguration. Therefore, the design of resilient Edge AI systems must carefully balance resilience goals with performance, energy consumption, and latency requirements.

Multi-objective optimization frameworks are increasingly used to balance trade-offs between performance, energy consumption, and resilience metrics in edge computing and AI systems [125, 126]. For example, edge schedulers can allocate compute resources and adjust model complexity in real time to maintain acceptable utility under constrained resources without compromising resilience.

## 4.7 Challenges and Future Directions

Despite advances in resilient mechanisms, several open challenges persist:

- *Scalable resilience:* Coordinating redundancy and anomaly detection across large, heterogeneous deployments.
- *Adversarial resilience:* Countering sophisticated attackers that exploit adaptive feedback loops.
- *Cross-layer integration:* Harmonizing resilience strategies across hardware, networking, and AI model layers.
- *Explainability:* Providing transparent accounts of degradation events and recovery processes to support accountability and regulatory compliance.

Future research will likely emphasize self-organizing resilience frameworks, trustworthy adaptive control, and context-aware resilience policies that dynamically adjust protection mechanisms according to risk assessments and operational criticality.

Cross-layer resilience mechanisms spanning device, network, and model layers are systematically summarized in Table 4.

**Table 4.** Cross-layer resilience mechanisms in distributed edge AI architectures

| Layer | Failure Type | Resilience Mechanism | Example Techniques |
|---|---|---|---|
| Hardware | Device failure | Redundancy | Replicated nodes |
| Network | Link disruption | Adaptive routing | Mesh networking |
| Learning | Byzantine behavior | Robust aggregation | Krum, median-based methods |
| Application | Service degradation | Graceful degradation | Load shedding |
| System-wide | Cascading failures | Orchestration policies | Distributed consensus |

# 5. Unified Trust Framework for Edge AI

The proliferation of intelligent edge devices has significantly reshaped the threat landscape of distributed artificial intelligence. Unlike centralized cloud AI, Edge AI systems operate across heterogeneous, resource-constrained, and often physically exposed environments, requiring trust to be established as a system-wide property rather than an isolated feature. In this context, trust emerges from the coordinated integration of security, privacy preservation, and resilience mechanisms across architectural layers. This section presents a unified trust framework that consolidates these dimensions into a cohesive architectural and operational model.

## 5.1 Foundations of Trust in Edge AI

Trustworthy Edge AI systems require guarantees across three interdependent dimensions: security, privacy, and resilience. Security focuses on safeguarding devices, communication channels, and models against unauthorized access, tampering, and adversarial manipulation. Privacy ensures protection of sensitive information through mechanisms such as data minimization, differential privacy, and secure aggregation. Resilience enables systems to maintain acceptable performance under faults, adversarial conditions, or resource variability.

Prior research underscores the need for integrated design approaches. Roman et al. highlight the expanded attack surface in edge environments and advocate security-by-design principles [11], while Shi et al. emphasize the need for coordinated safeguards across devices, networks, and cloud backends [1]. Complementary perspectives from trustworthy AI research extend these requirements to include robustness, fairness, and accountability. Liu et al. argue that such properties must be realized at the system level rather than in isolated components [127], providing a conceptual basis for unified trust architectures in Edge AI.

## 5.2 Integration of Security, Privacy, and Resilience in a Unified Model

In Edge AI systems, security, privacy, and resilience are inherently coupled and must be co-designed. Mechanisms that strengthen one dimension may introduce vulnerabilities in another. For example, federated learning preserves data locality but introduces risks such as model poisoning and backdoor attacks, while redundancy improves resilience but may expand the attack surface if not securely managed.

Federated learning exemplifies the need for integrated trust engineering. McMahan et al. introduced decentralized training with communication efficiency [8], followed by advances such as secure aggregation to protect model updates [29]. Comprehensive analyses by Kairouz et al. highlight challenges spanning adversarial robustness, privacy leakage, and system reliability [128].

A unified framework therefore integrates privacy-preserving communication, Byzantine-resilient aggregation, anomaly detection, and adaptive participation strategies. Approaches such as Byzantine-robust aggregation algorithms proposed by Blanchard et al. [129] demonstrate how resilience and security mechanisms can be embedded within privacy-aware training pipelines.

This integration becomes increasingly important in emerging 6G-enabled edge environments, where large-scale connectivity and ultra-low latency demand coordinated protections across device, communication, and computation layers [130].

## 5.3 Cross-Layer Protection: Device, Communication, and Algorithmic Levels

A unified trust framework requires coordinated defenses across device, communication, and algorithmic layers.

At the device layer, hardware roots of trust, secure boot, and trusted execution environments (TEEs) ensure system integrity and protect sensitive assets. Sabt et al. survey TEE technologies, including Intel SGX and ARM TrustZone, demonstrating how hardware-based isolation strengthens system security [131].

At the communication layer, encryption, authentication, and secure routing mechanisms protect data in transit. Lightweight cryptographic protocols and transport layer security (TLS)-based channels are particularly important in resource-constrained IoT environments. Alaba et al. emphasize the role of network-layer protections in mitigating edge-specific threats [132].

At the algorithmic layer, adversarial robustness techniques protect model integrity during training and inference. Madry et al. formalized adversarial training as a principled defense [26], while subsequent work extends robustness to distributed settings. Li et al. propose combining anomaly detection with robust aggregation to counter adversarial behavior in federated learning [133].

Effective trust emerges when these layers are integrated. For example, device-level attestation can inform higher-level decisions by excluding compromised nodes from collaborative learning, thereby preventing localized compromise from propagating system-wide.

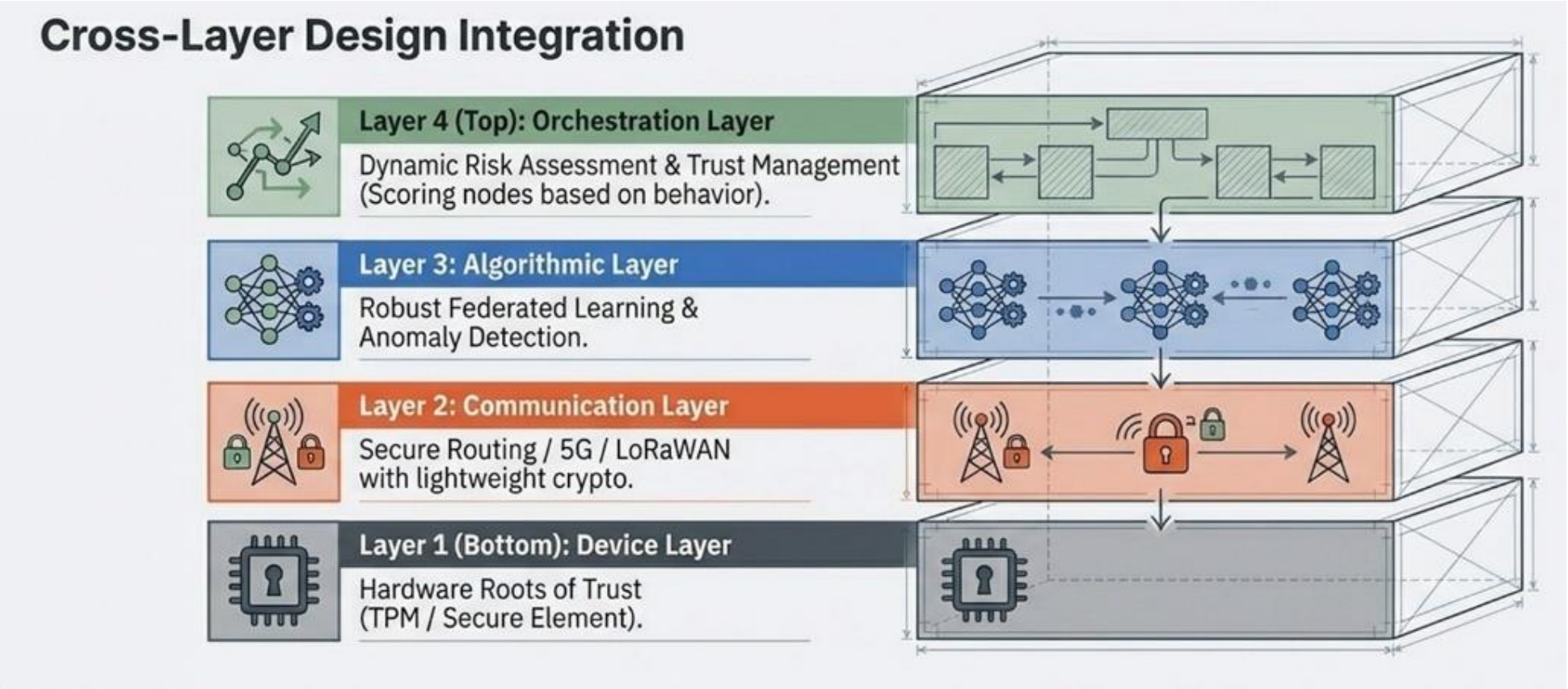

**Figure 5.** A cross-layer design integration framework for secure and robust distributed intelligence. The architecture illustrates a four-layer stack comprising: device layer, communication layer, algorithmic layer, and orchestration layer performing dynamic risk assessment and trust management.

Figure 5 highlights how trust established at the hardware level propagates through communication and learning layers, enabling coordinated enforcement and system-wide robustness.

The principal multi-objective design trade-offs in trustworthy Edge AI systems are summarized in Table 5.

**Table 5.** Design trade-offs in trustworthy edge AI systems

| Design Objective | Improves | Trade-Off With | Design Considerations |
|---|---|---|---|
| Strong encryption | Security, Privacy | Latency, Energy | Lightweight cryptography |
| Adversarial robustness | Resilience | Accuracy | Model complexity |
| Explainability | Transparency | Computation overhead | Selective explanation |
| Redundancy | Fault tolerance | Cost | Critical node prioritization |
| Continuous monitoring | Security | Bandwidth | Event-driven logging |

## 5.4 Dynamic Risk Assessment and Adaptive Trust Management

Static security policies are insufficient in dynamic edge environments. Adaptive trust management requires continuous evaluation of system behavior, contextual signals, and threat indicators to support real-time decision-making.

Trust modeling approaches in IoT systems provide useful foundations. Yan et al. survey context-aware and reputation-based trust mechanisms [134], which can be extended to Edge AI for dynamic trust scoring. In such systems, node behavior and communication patterns are continuously monitored, enabling early detection and isolation of anomalous entities [135, 136].

Anomaly detection plays a central role in this process. Chalapathy and Chawla review deep learning–based techniques for detecting irregular patterns in data and model behavior [94], supporting automated responses such as reconfiguration or retraining.

Emerging work explores reinforcement learning for adaptive defense strategies. Nguyen and Reddi discuss dynamic mitigation approaches under uncertainty [136], which can be applied to adjust system parameters such as sampling rates, aggregation rules, and security levels.

Adaptive privacy mechanisms further enhance trust. Differential privacy frameworks, as established by Dwork and Roth [36], can be extended to dynamically adjust privacy budgets based on contextual sensitivity and system requirements.

## 5.5 Conceptual Design of a Multi-Tiered Edge Trust Architecture

A unified trust architecture can be structured as a multi-tiered model comprising device-level trust anchors, edge-level orchestration, and cloud-level governance.

The device tier establishes hardware-rooted trust through secure elements and TEEs, enabling secure key management and attestation. The edge tier coordinates communication, performs anomaly detection, and manages federated learning with robust aggregation strategies. The cloud tier supports global model management, audit logging, and policy enforcement aligned with regulatory requirements.

Cloud–edge collaboration models, as described by Satyanarayanan [2], illustrate how computational responsibilities can be distributed to balance latency and scalability. Embedding trust mechanisms within each tier transforms this structure into a trust-aware processing pipeline.

Explainability and accountability further strengthen trust. Doshi-Velez and Kim emphasize the importance of interpretability in high-stakes AI systems [137], enabling diagnosis and remediation of anomalous outcomes.

From a governance perspective, distributed audit mechanisms such as blockchain-based logging have been explored to enhance accountability. Dorri et al. demonstrate how blockchain can support trust management in IoT systems [138], although practical deployment requires addressing latency and scalability constraints.

## 5.6 Design Strategies for End-to-End Trustworthy Edge AI

Designing trustworthy Edge AI systems requires balancing security, privacy, resilience, and resource efficiency. Defense-in-depth remains essential, but excessive redundancy or cryptographic overhead may exceed device constraints. Multi-objective optimization approaches can jointly address performance, energy consumption, and security requirements in edge scheduling [126].

Secure orchestration platforms must therefore incorporate context-aware policy engines capable of dynamically adapting to changing conditions. Zero-trust paradigms provide a useful foundation, emphasizing continuous verification and least-privilege access across distributed nodes.

A unified trust framework ultimately enables Edge AI systems to function as adaptive, context-aware ecosystems. By integrating hardware security, privacy-preserving computation, adversarial robustness, anomaly detection, and dynamic control within a cross-layer architecture, trust becomes an emergent system property.

Achieving this vision will require interdisciplinary advances in security engineering, privacy theory, robust machine learning, and distributed systems. Standardized evaluation frameworks, improved benchmarking of resilience and privacy leakage, and automated verification tools will be critical for validating trust guarantees in large-scale deployments.

# 6. Applications and Case Studies

Edge AI has moved beyond conceptual development and is now deployed across diverse real-world domains where security, privacy, and resilience are critical to operational success. Applications in smart cities, autonomous vehicles, healthcare wearables, and industrial IoT demonstrate how secure and resilient Edge AI enhances performance, reduces latency, preserves privacy, and maintains service continuity under system or network disruptions. This section presents representative applications and case studies illustrating how integrated trust frameworks function in practice and distills key lessons and design best practices from these deployments.

## 6.1 Smart Cities: Trusted Edge Nodes for Sensor Data Integrity

Smart city environments provide a compelling context for examining how trustworthy Edge AI principles are operationalized at scale. These deployments integrate diverse sensing modalities, real-time analytics, and distributed control systems, requiring coordinated mechanisms for security, privacy, and resilience. The following subsections illustrate how such principles are implemented in key urban applications, including intelligent traffic management and environmental monitoring.

### 6.1.1 Intelligent Traffic Management with Edge AI

In smart cities, Edge AI systems increasingly process large volumes of sensor data to enable real-time decision-making. Traffic management, infrastructure monitoring, environmental sensing, and emergency coordination benefit from placing intelligence close to data sources. A systematic survey on Edge AI for smart cities provides a comprehensive overview of layered architectures spanning sensing, learning models, and hardware infrastructure to support urban services in heterogeneous environments [139]. The survey synthesizes applications across transportation, healthcare, building management, and environmental monitoring, emphasizing that Edge AI reduces communication overhead and enables adaptive, resilient services even under intermittent connectivity.

In operational deployments, intelligent video analytics supports public safety by enabling edge nodes equipped with AI inference engines to process surveillance footage locally. These systems detect anomalies, identify unsafe behaviors, and support crowd management without transmitting raw video streams to centralized servers. Localized perception pipelines reduce latency, enhance privacy protection, and improve emergency responsiveness. Commercial traffic monitoring and adaptive signal control systems further demonstrate the practical viability of low-latency, data-driven urban intelligence [140].

A widely cited real-world implementation is the Hangzhou “City Brain” system [141]. The platform integrates thousands of traffic cameras and IoT sensors distributed throughout the city. Rather than transmitting raw video streams to centralized data centers, distributed edge processing clusters located near traffic intersections perform deep convolutional neural network inference for vehicle detection, traffic density estimation, violation detection, and incident recognition. The system adopts a hierarchical architecture: GPU-enabled edge servers conduct local feature extraction and event classification, while only compressed metadata and aggregated insights are transmitted to higher-level coordination nodes. This design reduces backhaul congestion and lowers end-to-end latency.

Resilience is achieved through multi-layer redundancy. Overlapping camera coverage ensures observational continuity, and computational workloads can be dynamically migrated between adjacent edge nodes in the event of failure. Model updates are deployed incrementally, allowing rapid rollback of faulty updates without disrupting traffic control. Security mechanisms include secure boot processes, encrypted inter-node communication, and identity-based authentication for sensor endpoints. Privacy protection is strengthened by processing identifiable video content locally and transmitting only anonymized metadata. Reported improvements in congestion reduction and emergency vehicle routing provide empirical validation of scalable, trusted edge intelligence. Figure 6 depicts intelligent traffic management in a smart city environment.

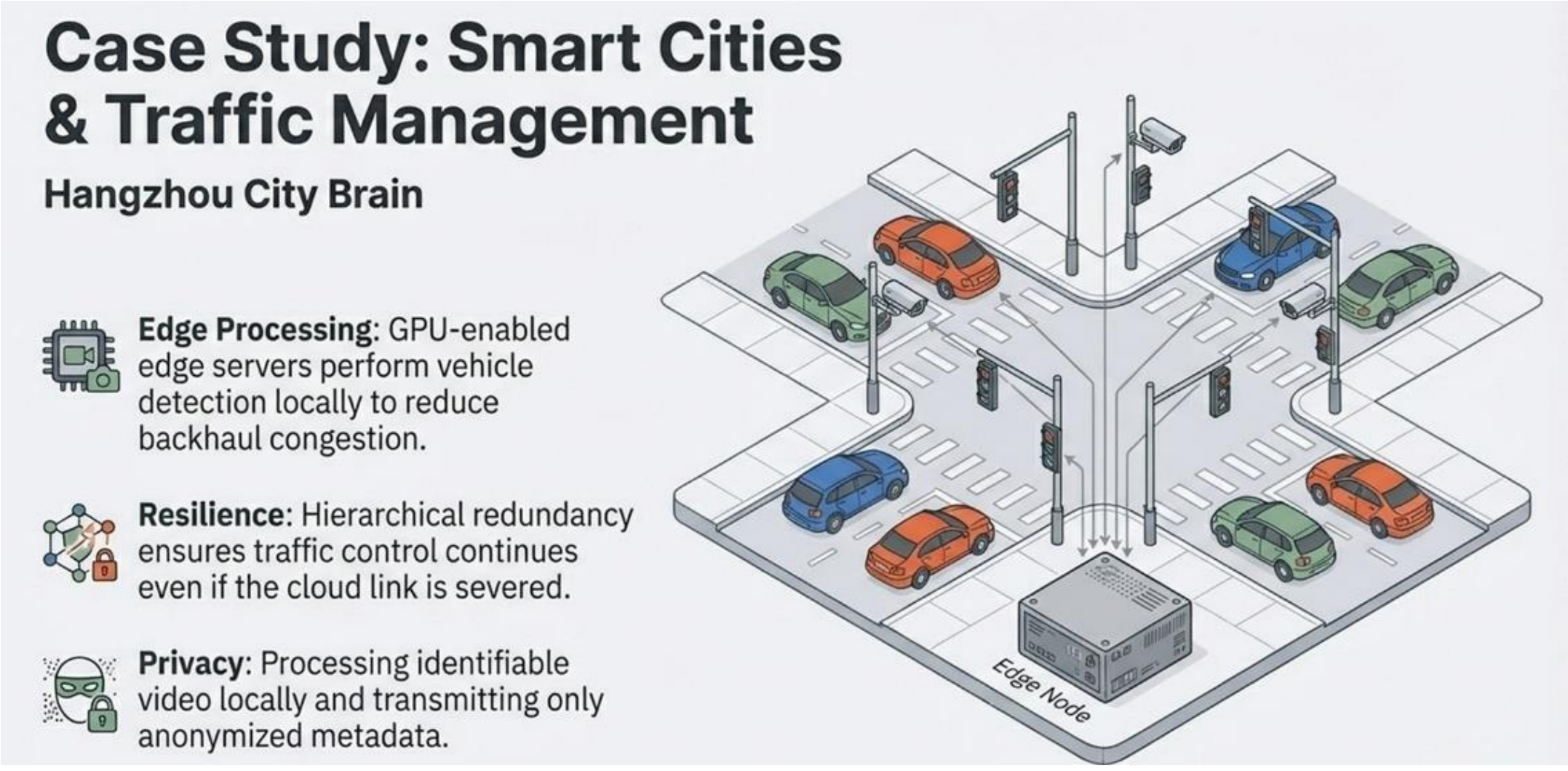

**Figure 6.** Case study of smart city-traffic management using edge intelligence (Hangzhou City Brain). The figure illustrates an edge-enabled urban traffic management framework where GPU-equipped edge nodes perform real-time vehicle detection and signal optimization at intersections. The architecture emphasizes (i) local edge processing to reduce backhaul congestion, (ii) hierarchical redundancy to ensure operational resilience during cloud disconnection, and (iii) privacy preservation through localized processing of identifiable video data with transmission of anonymized metadata only.

### 6.1.2 Case Study: Edge-Based Environmental Monitoring with Secure IoT Integration

Urban air quality monitoring increasingly leverages edge analytics to ensure responsiveness and data integrity. Low-cost sensors deployed across metropolitan regions are vulnerable to calibration drift and potential tampering. Contemporary deployments mitigate these risks through secure firmware, encrypted telemetry, and anomaly detection algorithms executed at edge gateways [142].

Edge nodes perform statistical filtering and outlier detection to identify abnormal readings indicative of sensor malfunction or spoofing. Cross-sensor validation enables spatial consistency checks among neighboring nodes. When discrepancies exceed predefined thresholds, automated recalibration workflows are triggered. Resilience strategies include redundant sensing clusters and distributed validation mechanisms. If individual nodes fail, neighboring sensors compensate using spatial interpolation models executed at the edge gateway. This deployment model demonstrates the convergence of authentication, encryption, redundancy, anomaly detection, and localized analytics within practical urban infrastructure.

## 6.2 Autonomous Vehicles: Fault-Tolerant Edge Perception and Cooperative Intelligence

Autonomous vehicle systems represent one of the most demanding applications of Edge AI, requiring ultra-low latency, high reliability, and robust decision-making under dynamic and safety-critical conditions. These systems integrate on-board perception, real-time inference, and cooperative intelligence across connected entities, necessitating tightly coupled mechanisms for resilience, security, and fault tolerance. The following case studies illustrate how redundant perception and collaborative sensing architectures enable dependable operation in both standalone and networked vehicular environments.

### 6.2.1 Case Study: Multi-Modal Redundant Perception in Autonomous Driving

Autonomous vehicles rely on on-board Edge AI for real-time perception. Modern platforms integrate cameras, radar, LiDAR, and ultrasonic sensors. Perception stacks typically execute on high-performance automotive edge processors such as NVIDIA DRIVE AGX platforms [143]. Resilience is achieved through both sensor and model redundancy. Independent neural networks process individual modalities, and fusion layers combine outputs using probabilistic frameworks or attention-based architectures. When one modality degrades due to adverse weather or adversarial perturbations, the fusion mechanism re-weights alternative inputs.

Security measures include cryptographically signed *over-the-air* (OTA) updates, hardware roots of trust, and intrusion detection within in-vehicle networks. Runtime monitoring identifies anomalous inference patterns potentially indicative of adversarial manipulation. Millisecond-level decision requirements necessitate fully localized inference, effectively transforming the vehicle into a self-contained edge data center.

### 6.2.2 Case Study: Cooperative Perception in Vehicle-to-Everything (V2X) Networks

Cooperative perception extends sensing capabilities beyond individual vehicles by leveraging roadside units (RSUs) and peer vehicles [144]. Edge servers at RSUs aggregate perception summaries from nearby vehicles and broadcast fused situational awareness maps. Distributed consensus mechanisms support resilience; neighboring nodes assume responsibility when an RSU fails. Vehicles retain fallback local perception in the event of communication loss.

Security mechanisms include certificate-based identity authentication and encrypted vehicle-to-everything (V2X) communication channels. Anomaly detection algorithms identify inconsistent or suspicious perception messages to prevent malicious data injection. Interruption-aware cooperative perception frameworks further enhance robustness by incorporating predictive recovery models and redundant contextual information to maintain accuracy under packet loss or communication disruptions [145].

Practical roadside perception systems employ optimized deep learning models (e.g., YOLOv8-Nano) deployed on energy-efficient embedded platforms such as NVIDIA Jetson devices. These systems detect traffic violations, generate standardized safety event messages, and support cooperative vehicle ecosystems with high accuracy and low power consumption [146]. The results illustrate how resource-efficient edge AI sustains real-time analytics and safety resilience.

## 6.3 Healthcare and Wearable Systems: Privacy-Preserving Edge Intelligence

Healthcare and wearable systems introduce stringent requirements for privacy, reliability, and real-time responsiveness, making them a critical domain for trustworthy Edge AI. Unlike many other applications, these systems must balance continuous monitoring with strict protection of sensitive patient data while ensuring dependable operation under variable connectivity and device constraints. The following case studies illustrate how federated learning, secure computation, and edge-based anomaly detection enable privacy-preserving and resilient healthcare intelligence.

### 6.3.1 Case Study: Federated Learning for Clinical Data Protection

Healthcare represents a domain where Edge AI deployment is expanding rapidly. Wearables, remote monitoring systems, and Internet of Medical Things (IoMT) platforms enable secure, low-latency health analytics. Federated learning combined with secure computation techniques allows patient data to remain local while contributing to collaborative model training. This paradigm has been applied in multi-institutional medical imaging, enabling collaborative tumor detection model development without sharing raw imaging datasets.

Decentralization enhances resilience: global training can proceed despite participant dropout, and local inference remains operational if aggregation servers become temporarily unreachable. Aminifar et al. propose a privacy-preserving edge federated learning architecture for mobile health and wearable IoT devices that incorporates cryptographic protections and resource-aware model training for applications such as epilepsy seizure detection [147].

### 6.3.2 Case Study: Continuous Vital Monitoring with Edge-based Anomaly Detection

Wearable devices continuously monitor physiological signals such as heart rate and blood oxygen levels. Embedded Edge AI algorithms, including lightweight CNN and LSTM models, detect anomalies such as arrhythmias in real time. A recent evaluation published in Scientific Reports describes a unified Edge-AI healthcare monitoring framework integrating federated learning, homomorphic encryption, dual wireless protocols (LoRaWAN and 5G), and lightweight blockchain-based audit logging [148].

In this architecture, CNN–LSTM models execute anomaly detection on embedded hardware with low latency. Homomorphic encryption protects model updates, and a Proof-of-Authority blockchain ensures secure audit trails. Reported performance metrics (approximately 91.9% accuracy) demonstrate the feasibility of secure, privacy-preserving, and resilient Edge AI for healthcare monitoring.

## 6.4 Industrial IoT: Redundant Edge Controllers and Predictive Maintenance

Industrial IoT environments represent mission-critical deployments of Edge AI, where uninterrupted operation, low latency, and system reliability are essential. These systems integrate distributed sensing, real-time analytics, and automated control loops, requiring robust mechanisms for fault tolerance, secure communication, and predictive maintenance. The following case studies illustrate how redundant edge controllers and distributed intelligence enable resilient and efficient industrial operations.

### 6.4.1 Case Study: Edge-Based Predictive Maintenance in Manufacturing

Industrial IoT environments illustrate how resilience and redundancy in edge controllers enhance operational continuity [149]. Manufacturing facilities deploy vibration and acoustic sensors on machinery, with edge controllers performing real-time fault detection using deep learning models. Instead of transmitting raw signals continuously to cloud servers, only anomaly alerts or compressed features are forwarded.

Resilience is achieved through redundant edge controllers configured in hot-standby mode, enabling seamless failover if a primary controller fails. Deterministic networking protocols support time-sensitive data delivery. Security measures include network segmentation, encrypted industrial communication protocols, and hardware-based device authentication. These deployments reduce downtime and demonstrate how edge intelligence sustains industrial operations.

### 6.4.2 Case Study: Cyber-Manufacturing with Distributed Edge Intelligence

Cyber-manufacturing integrates cyber-physical systems with edge analytics [150]. Edge nodes perform real-time process optimization by predicting deviations in production lines and issuing corrective adjustments autonomously. Hierarchical control structures enhance resilience: local controllers manage immediate corrections, while supervisory systems coordinate global optimization. Secure industrial communication protocols mitigate tampering risks, and the decentralized architecture reduces single points of failure and strengthens fault tolerance.

## 6.5 Cross-Domain Insights

Comparative analysis across domains reveals recurring design patterns in trustworthy Edge AI systems. Layered intelligence—combining near-device inference with federated or cloud-assisted aggregation—balances latency, privacy, and global learning objectives. For instance, healthcare deployments integrate real-time wearable anomaly detection with privacy-preserving federated updates to improve system-wide performance without exposing raw data.

Resilience is achieved through redundancy and adaptive mechanisms. Autonomous systems employ multi-modal sensing and fusion strategies to maintain performance under partial failure, while hybrid communication architectures (e.g., LoRaWAN and 5G) enable continuity through dynamic network adaptation.

Security and privacy must be co-designed with system performance. Techniques such as differential privacy, homomorphic encryption, and distributed audit mechanisms enhance confidentiality and accountability, but introduce computational and communication overheads that must be carefully managed.

Rigorous evaluation remains essential for real-world deployment. Systems must be assessed in terms of latency, energy efficiency, model accuracy, communication overhead, and regulatory compliance. Empirical validation on representative edge hardware platforms (e.g., Jetson Nano and ARM-based devices) provides critical insights into practical trade-offs.

Collectively, these observations highlight that trustworthy Edge AI emerges from integrated architectural design rather than isolated mechanisms, enabling reliable deployment across diverse and safety-critical application domains.

## 7. Emerging Trends in Secure and Trustworthy Edge AI

The evolution of Edge AI is increasingly shaped by the need to move beyond performance optimization toward systems that are inherently secure, resilient, interpretable, and compliant with emerging regulatory and ethical requirements. As Edge AI becomes integral to critical domains, future research is focusing on advancing foundational capabilities while ensuring their practical deployability in resource-constrained and distributed environments.

A key technical direction is the development of *lightweight cryptographic mechanisms* tailored for constrained edge devices. Traditional cryptographic schemes often impose prohibitive computational overhead, motivating the adoption of efficient alternatives. The National Institute of Standards and Technology (NIST) standardized the Ascon algorithm for lightweight authenticated encryption in 2023, following rigorous evaluation of efficiency and security properties [151, 152]. Such primitives enable secure communication and data integrity with minimal resource consumption, making them suitable for edge sensors and distributed learning systems.

Efficient public-key cryptography is also evolving to meet edge requirements. Elliptic-curve–based schemes, including Curve25519, provide strong security with reduced computational overhead compared to conventional approaches. Bernstein and Lange demonstrate that optimized elliptic-curve implementations can achieve high performance even on constrained hardware [153], supporting applications such as secure boot, firmware updates, and remote attestation.

Hardware-assisted security continues to play a critical role in establishing trust anchors. Trusted Execution Environments (TEEs) provide isolated execution for sensitive computations and enable remote attestation of device integrity. As noted by Roman et al. [11], combining hardware roots of trust with system-level protections is essential for secure operation in distributed edge environments, particularly in collaborative settings where node integrity must be verified.

Post-quantum cryptography introduces new design considerations. Hybrid schemes combining lightweight symmetric encryption with quantum-resistant key exchange mechanisms are being explored to ensure forward security while maintaining feasibility on constrained devices [154].

In parallel, *distributed learning paradigms* continue to evolve. Federated learning remains central to privacy-preserving Edge AI, enabling decentralized training without raw data sharing [8]. Current research focuses on strengthening robustness and trust in such systems. Byzantine-resilient aggregation methods, such as Krum [129], provide defenses against malicious updates, while secure aggregation protocols [29] protect update confidentiality. Surveys by Li et al. highlight ongoing challenges, including heterogeneity, communication efficiency, and adversarial resilience [38]. Emerging approaches incorporate adaptive mechanisms such as trust-weighted aggregation and anomaly-based client selection to improve robustness in dynamic environments.

Explainable AI (XAI) is gaining importance as Edge AI systems are deployed in high-stakes scenarios. Interpretability techniques, including LIME [155] and SHAP [156], support transparency and accountability, although their computational overhead poses challenges for edge deployment. Research is therefore shifting toward lightweight and inherently interpretable models that integrate explanation mechanisms directly into system design. As emphasized by Doshi-Velez and Kim [137], achieving a balance between interpretability and performance remains a central challenge.

Regulatory developments are simultaneously shaping system design requirements. The European Union's Artificial Intelligence Act (2024) establishes a risk-based framework for AI governance, imposing strict requirements on high-risk applications [157]. For Edge AI, compliance necessitates transparent data handling, continuous monitoring, and robust risk management practices.

The alignment between technical safeguards and governance requirements for trustworthy Edge AI deployment is summarized in Table 6.

**Table 6.** Governance and regulatory alignment for trustworthy Edge AI deployment

| Governance Dimension | Technical Mechanisms | Documentation Required | Stakeholders |
|---|---|---|---|
| Data protection | Differential privacy, encryption | Data impact assessments | Operators |
| Transparency | Explainable AI | Audit logs | Regulators |
| Accountability | Identity attestation | Provenance records | Vendors |
| Safety | Redundancy, formal verification | Risk assessment reports | Certification Bodies |
| Lifecycle management | Secure decommissioning | Disposal records | Manufacturers |

Complementary frameworks further reinforce governance practices. The NIST AI Risk Management Framework (AI RMF 1.0) provides structured guidance for managing AI risks across system lifecycles [158], while ISO/IEC 23894:2023 establishes international standards linking technical robustness with organizational accountability [159]. These frameworks are particularly relevant for Edge AI deployments in regulated sectors.

Ethical considerations, including fairness and bias mitigation, are also becoming central. Distributed learning systems may amplify data imbalance across populations, necessitating fairness-aware optimization and auditing mechanisms. Addressing these concerns is critical for ensuring equitable and trustworthy system behavior.

Sustainability is emerging as an additional design dimension. While edge computing reduces reliance on centralized infrastructure, it increases distributed hardware usage. Techniques such as model pruning, quantization, and knowledge distillation reduce energy consumption and support environmentally responsible deployment.

Finally, adaptive trust management represents a promising research direction. Reinforcement learning–based approaches enable dynamic system responses to evolving threats. Nguyen and Reddi [136] highlight how such techniques can support automated mitigation strategies, suggesting their potential for real-time orchestration of security, privacy, and performance trade-offs in Edge AI systems.

In summary, future Edge AI systems will depend on the integration of efficient cryptography, robust distributed learning, explainable models, regulatory compliance, ethical safeguards, and sustainability considerations. The convergence of these directions will determine the extent to which Edge AI can serve as a reliable and trustworthy foundation for next-generation intelligent infrastructure.

## 8. Conclusion

The evolution of Edge AI marks a fundamental transition from centralized cloud computing to distributed intelligence across heterogeneous, resource-constrained, and geographically dispersed environments. This shift not only transforms system architectures but also redefines how trust, reliability, and accountability must be engineered. As discussed throughout this chapter, trustworthy Edge AI cannot be achieved through isolated mechanisms; it requires a holistic, cross-layer design approach in which security, privacy, and resilience are treated as foundational system properties.

At the system level, the distributed topology of Edge AI enables low-latency inference and real-time responsiveness but simultaneously expands the attack surface. Edge devices operate in physically exposed and dynamically changing environments, making traditional perimeter-based security models insufficient. Instead, robust trust must be established through zero-trust principles, hardware-rooted identity, secure boot mechanisms, encrypted communication, and continuous attestation to ensure verifiable operation across all nodes.

Privacy preservation remains equally critical, as Edge AI systems process sensitive data such as biometric signals, location information, and industrial telemetry. Decentralized learning paradigms, including federated learning, secure aggregation, and differential privacy, enable collaborative intelligence without centralizing raw data. However, these approaches introduce challenges related to communication overhead, system heterogeneity, and adversarial vulnerabilities, necessitating integration with cryptographic safeguards, secure hardware, and regulatory compliance frameworks.

Resilience forms the third pillar of trustworthy edge intelligence, ensuring sustained operation under failures, adversarial conditions, and resource variability. Mechanisms such as redundancy, anomaly detection, adaptive control, and robust distributed optimization enable systems to maintain acceptable performance despite disruptions. Importantly, resilience extends beyond fault tolerance to include robustness against adversarial manipulation and coordinated failures in distributed environments.

A central insight is the interdependence of security, privacy, and resilience. Enhancing one dimension often introduces trade-offs in others, such as increased latency, energy consumption, or reduced model accuracy. Addressing these trade-offs requires multi-objective optimization strategies that balance performance, efficiency, and trust guarantees within constrained environments.

Equally important is cross-layer integration. Trust cannot be ensured by isolated components; it must emerge from coordinated interactions across hardware, communication, algorithmic, and orchestration layers. Such integration enables adaptive and self-healing systems, where anomalies detected at one layer can trigger corrective actions across others, enhancing overall system robustness.

As Edge AI continues to converge with domains such as smart cities, autonomous systems, industrial automation, and healthcare, the need for comprehensive trust engineering becomes increasingly critical. Regulatory frameworks, risk management standards, and governance mechanisms further reinforce the importance of transparency, accountability, and compliance in distributed AI systems. At the same time, emerging trends, including lightweight cryptography, explainable AI, adaptive trust management, and sustainability-driven design, are shaping the next generation of edge intelligence.

Looking forward, Edge AI systems are expected to evolve into adaptive, context-aware ecosystems capable of dynamically responding to environmental changes and threat conditions. Achieving this vision will require advances in distributed learning, secure computation, interoperability, and energy-efficient design, supported by interdisciplinary collaboration across machine learning, security, systems engineering, and policy domains.

Ultimately, the promise of Edge AI lies in delivering scalable, real-time intelligence while upholding security, privacy, and societal trust. By adopting holistic design principles and addressing emerging challenges proactively, Edge AI can serve as a dependable foundation for next-generation intelligent infrastructures across diverse and safety-critical applications.

## Glossary

*Adversarial Attack:* A technique where inputs are intentionally modified to deceive machine learning models into making incorrect predictions.

*Adversarial Robustness:* The ability of a machine learning model to resist adversarial inputs.

*Anomaly Detection:* Techniques used to identify unusual patterns that may indicate faults or attacks.

*Byzantine Fault:* A failure mode where components provide incorrect or malicious information to the system.

*Differential Privacy (DP)*: A mathematical framework that protects individual data by adding controlled noise to computations, ensuring that outputs do not reveal sensitive information.

*Distributed Intelligence*: A system architecture in which computational and decision-making capabilities are spread across multiple nodes instead of a central location.

*Edge AI:* A paradigm where artificial intelligence models are deployed and executed on devices close to the data source rather than centralized cloud servers.

*Fault Tolerance:* The capability of a system to continue functioning even when some components fail.

*Federated Learning (FL)*: A decentralized machine learning approach where multiple devices collaboratively train a model without sharing raw data.

*Fully Homomorphic Encryption (FHE)*: A form of homomorphic encryption that allows arbitrary computations (both addition and multiplication) to be performed directly on encrypted data without requiring decryption. The result of these computations, when decrypted, matches exactly with the outcome of performing the same operations on the original plaintext data. FHE enables secure outsourced computation and privacy-preserving machine learning but typically incurs high computational and latency overhead, making its deployment challenging in resource-constrained edge environments.

*Hardware Root of Trust:* A secure hardware component that serves as the foundation for system security (e.g., secure boot, key storage).

*Homomorphic Encryption (HE):* A cryptographic technique that allows computations to be performed directly on encrypted data without decryption.

*Membership Inference Attack:* An attack that determines whether a specific data sample was used in training a model.

*Model Inversion Attack:* A technique used to reconstruct input data from a trained model.

*Model Poisoning:* A type of attack in which malicious participants manipulate training data or updates to corrupt a machine learning model.

*Resilience:* The ability of a system to maintain acceptable performance under failures, attacks, or adverse conditions.

*Secure Aggregation:* A cryptographic method that allows aggregation of model updates while keeping individual contributions private.

*Secure Boot:* A process ensuring that a device boots only trusted and verified software.

*Secure Multiparty Computation (SMC):* A method enabling multiple parties to jointly compute a function while keeping their inputs private.

*Trusted Execution Environment (TEE):* A secure area within a processor that ensures sensitive code and data are protected from unauthorized access.

*Zero-Trust Architecture:* A security model that assumes no implicit trust and requires continuous verification of all entities in a system.